\documentclass[runningheads]{llncs}
\usepackage[T1]{fontenc}
\usepackage{graphicx}
\usepackage{tikz}
\usetikzlibrary{automata,arrows,positioning}
\usepackage{smartdiagram}
\usesmartdiagramlibrary{additions}

\usepackage{hyperref}
\usepackage{color}

\begin{document}
\title{Enabling Portability and Reusability of\\ Open Science Infrastructures}
%
%
\author{Giuseppe Grieco\orcidID{0000-0001-5439-4576} \and
Ivan Heibi\orcidID{0000-0001-5366-5194},
Arcangelo Massari\orcidID{0000-0002-8420-0696}, Arianna Moretti\orcidID{0000-0001-5486-7070} \and Silvio Peroni\orcidID{0000-0003-0530-4305}}
\titlerunning{Enabling Portability and Reusability of Open Science Infrastructures}
\authorrunning{Grieco et al.}
%
\institute{Research Centre for Open Scholarly Metadata, Department of Classical Philology and Italian Studies, University of Bologna, Bologna, Italy \email{g.grieco1997@gmail.com,ivan.heibi2@unibo.it,arcangelo.massari@unibo.it,\\arianna.moretti4@unibo.it,silvio.peroni@unibo.it}}
\maketitle              
\begin{abstract}
This paper presents a methodology for designing a containerized and distributed open science infrastructure to simplify its reusability, replicability, and portability in different environments. The methodology is depicted in a step-by-step schema based on four main phases: (1) Analysis, (2) Design, (3) Definition, and (4) Managing and provisioning. We accompany the description of each step with existing technologies and concrete examples of application.

\end{abstract}

\section{Introduction}
Open Science Infrastructures (OSInfras) are resources and services that the scholarly ecosystem depends upon to foster research and ``to support open science and serve the needs of different communities'' \cite{unesco_unesco_2021}. According to a survey published in 2020 \cite{ficarra_victoria_2020_4159838}, there are 120 OSInfras in Europe, heterogeneous by domain and objectives. In recent years, several founders -- including the European Union with its financial support towards building the European Open Science Cloud (EOSC, \url{https://eosc-portal.eu/about/eosc}) -- and institutions, such as UNESCO with its Open Science recommendations \cite{unesco_unesco_2021}, have strongly emphasised how the survival of OSInfras is crucial for enabling Open (i.e. good) Science.

An OSInfra is made by several complementary pillars that concern (a) technological aspects (i.e. ``software, hardware, and technical services'' \cite{lin_trust_2020}), (b) social (i.e. the people behind the infrastructures) and (c) economic endeavours (i.e. their sustainability in the long term). Several guidelines, such as \cite{Bilder2020-nh} \cite{skinner_values_2020} \cite{confederation_of_open_access_repositories_good_2019}, have been published to help the scholarly community  running, monitoring, and maintaining OSInfras in all these aspects.

Focusing on technological concerns, several of these guidelines agree on adopting open source software for running OSInfras' services. Indeed, both the \emph{Principles for Open Scholarly Infrastructures} \cite{Bilder2020-nh} and another recent report by the Knowledge Future Group about the values and principles for an OSInfra \cite{skinner_values_2020} mention using open software, technologies, standards, and protocols. Such principles are essential for ensuring that the OSInfra can be reusable and portable into new organisations if the original maintainer is not capable anymore of handling it. These aspects concerning the reusability (in the FAIR sense \cite{Wilkinson2016-te} \cite{Chue_Hong2021-wv} \cite{HasselbringCarrHettrickPackerTiropanis+2020+39+47} \cite{Lamprecht2020-fb}) and portability of OSInfras are crucial values to guarantee. Indeed, in \cite{skinner_values_2020}, the authors stress that an OSInfra should enable and encourage the reuse of code, and ensure the portability and durability of the content (including software and services) that it hosts. Others explicitly ask to enable easy migration of such content to another platform if needed \cite{confederation_of_open_access_repositories_good_2019}, guaranteeing that all the ongoing assets can be ``archived and preserved when passed to a successor organisation'' \cite{Bilder2020-nh}.

An OSInfra is a complex system providing several services that can be either tied up into a monolithic container or distributed in distinct locations federated via APIs, and even if the software for replicating the OSInfra is released with open source licenses, this is not enough to guarantee reusability, portability, and redistribution of the OSInfra. Indeed, specific documentation and tools should be considered to allow an easy reuse and deployment in a different environment.

In this paper, to address the issues mentioned above, we present a methodology in four steps that proposes the adoption of existing technologies to enable the isolation, federation and distribution of the services of individual OSInfras to simplify their reusability, replicability and portability. The solution we propose is tied with the \emph{infrastructure-as-code} (IaC) practice \cite{7965401}, where we use a standard language to design an infrastructure, including aspects related to scripting, automation, configuration, models, required dependencies, and parameters. This approach is combined with methods based on containers for separation and isolation of services to foster a more interoperable application packaging \cite{7092949}, platform-as-a-service (PaaS) runtimes \cite{7036275}, and a better scalability and reliability \cite{7742215} of services, so that the software modification could be done directly on the desired service without impacting the other ones provided by the OSInfra \cite{7742215}. 

All the steps of the methodology, introduced in Section \ref{methodology}, are accompanied by examples of (future) applications on OpenCitations (\url{https://opencitations.net}) \cite{10.1162/qss_a_00023}, i.e. an existing OSInfra dedicated to the publication of open bibliographic metadata and citation data. Finally, in Section \ref{conclusions}, we conclude the paper sketching out some future works.

\section{Methodology}\label{methodology}
\begin{figure}
    \centering
    \includegraphics[width=90mm,scale=1]{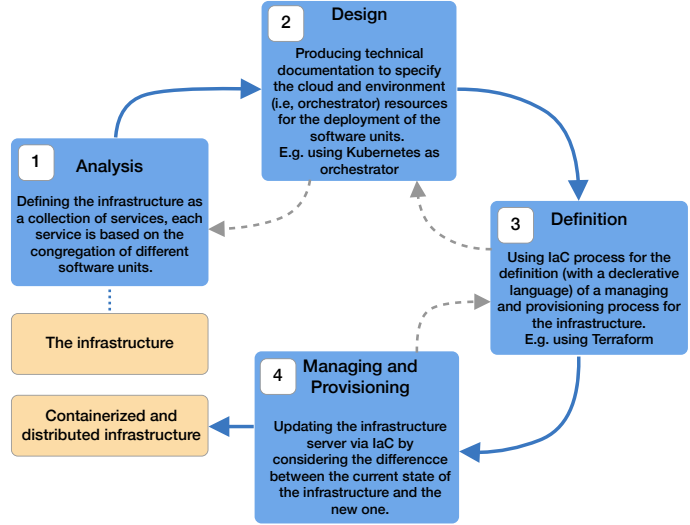}
    \caption{The workflow summarising the steps of the proposed methodology.}
    \label{fig:methodology}
\end{figure}

As summarised in Fig. \ref{fig:methodology}, our methodology is based on four steps: (1) Analysis, (2) Design, (3) Definition, and (4) Managing and provisioning, that are detailed in the following subsections. The workflow of the methodology is bidirectional: in clockwise, the output of each step becomes the input of the following one; in counterclockwise, it enables a backward step (an explanation on when it is needed is discussed in the following subsections) to re-process and refine the output returned previously. In addition, the methodology is not entirely connected in a closed circle since the output of step 4 \emph{is not} given as input to step 1 -- and, thus, any counterclockwise move from step 1 to step 4 is prohibited. 

\subsection{Analysis}
The aim of this step is to define a new organization to the infrastructure as a collection of separated services, each of them defined as a composition of different software units. This step is structured in two sub-steps. 

\textbf{First sub-step: software units.} We analyze the software units (e.g. specific libraries and applications) used by the infrastructure. This process should be done with the calculation of a trade-off between \emph{decoupling} and \emph{cohesion} \cite{Candela2016-sh}, that are crucial aspects to consider for determining how well components communicate with each other and with the end-user. Decoupling avoids situations where highly coupled components cause intensive intra-infrastructure traffic and are logically codependent. Instead, when components are highly cohesive, managing the overall load balancing is challenging since it could be hard to isolate the components for which more resources are needed. In addition, a wise choice of the trade-off between these two aspects permits the integration of other third party components (e.g. software) inside the infrastructure. This aspect is particularly relevant to support a federated infrastructure.

If any problem arises during the definition of the software units, then a document should be produced highlighting how to improve the cohesion and decoupling trade-off with respect to infrastructure requirements. Issues detected in this phase do not concern system efficiency but rather the evaluation of relevant factors that might impact the logical design of a distributed infrastructure.

For example, OpenCitations (as of 5 June 2022) handles everything through one service, which is highly cohesive since it is the main hub in charge of several other sub-services, such as the website, the APIs, and the access to the stored collections. Therefore, in this case, a document should be produced to guide OpenCitations' software engineers to improve such a huge cohesiveness before moving to the next sub-step. 

\textbf{Second sub-step: services.} Once the trade-off between cohesion and decoupling is verified, the software units are organized into services to isolate the work of the different parts of the infrastructure. Each single service collects the software units which are logically related and relevant to its functionality. Considering the current main OpenCitations service, it should be split into several other services (that use the software units defined in the previous step). For instance, such new services should include the OpenCitations \textit{website}, the \textit{REST APIs}, and the \textit{database access}. 

In case we are iterating again the methodology to build a new version of the infrastructure to include new services, this step processes only the new additional services and extends the previous documentation reporting the services and the software units managed.

\subsection{Design}
In this step, we generate a technical documentation that specifies the resources needed by the software units composing the services provided by the infrastructure. The documentation describes the resources to be created in the cloud, e.g. virtual machines with specific computational and storage capacities provided to cloud users by particular cloud provider such as the Amazon Web Services, and in an environment managed by the \emph{orchestrator}, i.e. a software agent that defines how to select, deploy, monitor, and dynamically control the configuration of multi-container packaged applications \cite{puliafito_container_2019}. 
A popular orchestrator tool is Kubernetes \cite{hightower_kubernetes_2017}. In Kubernetes, an object is a \emph{record of intent}: once it is created, Kubernetes constantly works to ensure that such object keeps working. In other words, creating an object tells Kubernetes how we want its workload to be handled in terms of usage of resources, including hardware resources and behaving policies (e.g, upgrades and fault-tolerance).

In OpenCitations, Kubernetes should be used to specify a \emph{pod} (the smallest deployable unit of computing in Kubernetes) and its deployment specifications (e.g. load balancing) for each individual service, e.g. the \textit{website} and \textit{REST APIs}. Each pod groups all the containers needed to run a corresponding software unit needed by a particular service -- for instance, in case of the OpenCitations \textit{website}, we might use a pod for the database used for \textit{authentication} and another for the \textit{HTTP web server}. It should also be necessary to specify the hardware resources and network requirements for each pod, e.g. deciding to accept or not incoming requests external to the Kubernetes cluster -- For example, for the \textit{website}, we can grant to accept HTTP GET requests since it needs to be exposed externally.

The output of this step is a documentation which groups the containers and the cloud resources needed by each service, and defines the overall design of the infrastructure. Of course, we can go back to the previous step in case we find out that the services partitioning is not satisfying/correct, e.g., due to the inclusion of unrelated software units or if we think there are services that incorporate too many software units. 

\subsection{Definition}
In this step, we define the design of the infrastructure using infrastructure-as-a-code (IaC) – a process for managing and provisioning an infrastructure by defining it through declarative language instead of using classical tools based on configuration files, CLI, and control panels \cite{7819402}. In IaC, the declarative language specifies the desired state of the infrastructure, and lets the actions to achieve it be automatically inferred. One of the possible tools to adopt for declarative IaC is Terraform \cite{brikman_terraform_2017}, a software for defining, launching, and managing IaC across a variety of cloud and virtualization platforms.

Using IaC gives us several advantages. It enables the unification of all resource definitions using a standard language, thus facilitating both maintenance and understanding by external adopters. In addition, specifying all the parameters for deployment in appropriate configuration files simplifies the infrastructure migration process, which is of particular relevance for supporting portability of the OSInfra, in case the organization decides to no longer maintain its services. Indeed, these aspects of IaC favor the organizations willing to reuse the infrastructure's services and preserve its heritage \cite{Bilder2020-nh}, ensure the development of a highly maintainable and sustainable software product \cite{Chue_Hong2021-wv}, and foster reproducibility and reusability by facilitating OSInfra understanding and trust \cite{Wilkinson2016-te}.

In this step, the resources (i.e. cloud and environment ones) are coded following the requirements established during the design phase. It might be necessary to return to the design phase if we realize that the infrastructure model does not provide sufficient detail on the resources needed, or in case some necessary resources are not included. In OpenCitations, we can use Terraform to declare the resources needed by each of the services following the documentation provided in the previous phase -- for instance, the pods and the network configurations needed by OpenCitations \textit{website} service.

\subsection{Managing and provisioning}
This is the final step of the methodology, it takes in input the state of the infrastructure defined via IaC and updates the remote state of the server with respect to such definitions. This operation is accomplished again via IaC. Depending on the IaC technology used, the state of the infrastructure could be updated using two different strategies: push strategy — the state is sent to the recipient servers, or pull strategy — the state is pulled by the recipient servers. In case this is not the first iteration of the methodology, the state of the infrastructure is updated considering the delta between the current state of the infrastructure and the new one.

To evaluate the result of this phase and decide whether to go back to the previous step or not, benchmarks on the infrastructure are needed to assess the infrastructure efficiency from a technical point of view. It is worth mentioning that it is difficult to obtain the optimal infrastructure after one iteration, therefore it is highly expected to step backwards to previous steps and refine the results until we finally obtain the \emph{desired} output.    

The term \emph{desired} is deliberately ambiguous, because the constraints might not be purely technical, e.g. the number of users to be supported or the financial limitations to respect. Therefore, a benchmark strategy for this step should test the infrastructure considering all these constraints.

In OpenCitations, concerning this step, we should design benchmarks for all the services, e.g. the \textit{website}, the \textit{REST APIs} and the \textit{database access}, for instance through the application of massive stress tests on the services.

\section{Discussion and conclusions}\label{conclusions}
Re-engineering an OSInfra from one single monolithic to a containerized and distributed model increases the scalability and reliability of its services. A continuous benchmark analysis of the system is essential to achieve the desired result, since the performances of the infrastructure components may vary with a large degree of unpredictability considering the new factors involved in the new distributed model.

One of the crucial aspects of this methodology concerns the use of IaC as a mean to promote the reproducibility and reusability of the infrastructure. IaC has been applied in literature for the research software. However, in this paper, we have abstracted this approach to involve the technical organisation of an OSInfra.  

It is necessary that the implementation of each phase of the methodology is followed by software engineers and software developers. In addition, from an administrative point of view, the maintenance and management of this architectural model requires a continuous configuration, monitoring, and optimization of the components composing the infrastructure. 

Finally, the methodology has been designed to be flexible and adaptable to specific use cases. Therefore, it is possible to integrate additional in-between sub-steps to address specific requirements, e.g. to refine the output of a step or to add other technical output required by a next step. Our upcoming plan is to apply the methodology to re-engineer the current OpenCitations technical infrastructure.

\subsubsection{Acknowledgements.} The work has been partially funded by the European Union's Horizon 2020 research and innovation program under grant agreement No 101017452 (OpenAIRE-Nexus).

 
%
%
%
%
 \bibliographystyle{splncs04}
 \bibliography{references}

\begin{thebibliography}{10}
\providecommand{\url}[1]{\texttt{#1}}
\providecommand{\urlprefix}{URL }
\providecommand{\doi}[1]{https://doi.org/#1}

\bibitem{7965401}
Artac, M., Borovssak, T., Di~Nitto, E., Guerriero, M., Tamburri, D.A.: Devops:
  Introducing infrastructure-as-code. In: 2017 IEEE/ACM 39th International
  Conference on Software Engineering Companion (ICSE-C). pp. 497--498 (2017).
  \doi{10.1109/ICSE-C.2017.162}

\bibitem{7036275}
Bernstein, D.: Containers and cloud: From lxc to docker to kubernetes. IEEE
  Cloud Computing  \textbf{1}(3),  81--84 (2014). \doi{10.1109/MCC.2014.51}

\bibitem{Bilder2020-nh}
Bilder, G., Lin, J., Neylon, C.: The principles of open scholarly
  infrastructure (2020). \doi{10.24343/C34W2H}

\bibitem{brikman_terraform_2017}
Brikman, Y.: Terraform: {Up} and {Running} {Writing} {Infrastructure} as
  {Code}. O'Reilly Media, Inc., 1st edn. (2017)

\bibitem{Candela2016-sh}
Candela, I., Bavota, G., Russo, B., Oliveto, R.: Using cohesion and coupling
  for software remodularization. ACM Trans. Softw. Eng. Methodol.
  \textbf{25}(3),  1--28 (Aug 2016). \doi{10.1145/2928268}

\bibitem{puliafito_container_2019}
Casalicchio, E.: Container {Orchestration}: {A} {Survey}. In: Puliafito, A.,
  Trivedi, K.S. (eds.) Systems {Modeling}: {Methodologies} and {Tools}, pp.
  221--235. Springer International Publishing, Cham (2019).
  \doi{10.1007/978-3-319-92378-9_14},
  \url{http://link.springer.com/10.1007/978-3-319-92378-9_14}, series Title:
  EAI/Springer Innovations in Communication and Computing

\bibitem{Chue_Hong2021-wv}
Chue~Hong, N.P., Katz, D.S., Barker, M., Lamprecht, A.L., Martinez, C.,
  Psomopoulos, F.E., Harrow, J., Castro, L.J., Gruenpeter, M., Martinez, P.A.,
  Honeyman, T.: {FAIR} {Principles} for {Research} {Software} ({FAIR4RS}
  {Principles}). Recommendations with {RDA} {Endorsement} in {Process},
  Research Data Alliance (2022), \url{https://doi.org/10.15497/RDA00068}

\bibitem{confederation_of_open_access_repositories_good_2019}
{Confederation Of Open Access Repositories}, {SPARC*}: Good {Practice}
  {Principles} for {Scholarly} {Communication} {Services}. Tech. rep.,
  {Confederation Of Open Access Repositories} and {SPARC*} (2019),
  \url{https://sparcopen.org/our-work/good-practice-principles-for-scholarly-communication-services/}

\bibitem{7742215}
Fazio, M., Celesti, A., Ranjan, R., Liu, C., Chen, L., Villari, M.: Open issues
  in scheduling microservices in the cloud. IEEE Cloud Computing
  \textbf{3}(5),  81--88 (2016). \doi{10.1109/MCC.2016.112}

\bibitem{ficarra_victoria_2020_4159838}
Ficarra, V., Fosci, M., Chiarelli, A., Kramer, B., Proudman, V.: {Scoping the
  Open Science Infrastructure Landscape in Europe} (Oct 2020).
  \doi{10.5281/zenodo.4159838}, \url{https://doi.org/10.5281/zenodo.4159838}

\bibitem{HasselbringCarrHettrickPackerTiropanis+2020+39+47}
Hasselbring, W., Carr, L., Hettrick, S., Packer, H., Tiropanis, T.: From fair
  research data toward fair and open research software. it - Information
  Technology  \textbf{62}(1),  39--47 (2020). \doi{doi:10.1515/itit-2019-0040},
  \url{https://doi.org/10.1515/itit-2019-0040}

\bibitem{hightower_kubernetes_2017}
Hightower, K., Burns, B., Beda, J.: Kubernetes: {Up} and {Running} {Dive} into
  the {Future} of {Infrastructure}. O'Reilly Media, Inc., 1st edn. (2017)

\bibitem{7819402}
Johann, S.: Kief morris on infrastructure as code. IEEE Software
  \textbf{34}(1),  117--120 (2017). \doi{10.1109/MS.2017.13}

\bibitem{Lamprecht2020-fb}
Lamprecht, A.L., Garcia, L., Kuzak, M., Martinez, C., Arcila, R., Martin
  Del~Pico, E., Dominguez Del~Angel, V., van~de Sandt, S., Ison, J., Martinez,
  P.A., McQuilton, P., Valencia, A., Harrow, J., Psomopoulos, F., Gelpi, J.L.,
  Chue~Hong, N., Goble, C., Capella-Gutierrez, S.: Towards {FAIR} principles
  for research software. Data sci.  \textbf{3}(1),  37--59 (Jun 2020).
  \doi{10.3233/DS-190026}

\bibitem{lin_trust_2020}
Lin, D., Crabtree, J., Dillo, I., Downs, R.R., Edmunds, R., Giaretta, D.,
  De~Giusti, M., L’Hours, H., Hugo, W., Jenkyns, R., Khodiyar, V., Martone,
  M.E., Mokrane, M., Navale, V., Petters, J., Sierman, B., Sokolova, D.V.,
  Stockhause, M., Westbrook, J.: The {TRUST} {Principles} for digital
  repositories. Scientific Data  \textbf{7}(1), ~144 (Dec 2020).
  \doi{10.1038/s41597-020-0486-7},
  \url{http://www.nature.com/articles/s41597-020-0486-7}

\bibitem{7092949}
Morabito, R., Kjällman, J., Komu, M.: Hypervisors vs. lightweight
  virtualization: A performance comparison. In: 2015 IEEE International
  Conference on Cloud Engineering. pp. 386--393 (2015).
  \doi{10.1109/IC2E.2015.74}

\bibitem{10.1162/qss_a_00023}
Peroni, S., Shotton, D.: {OpenCitations, an infrastructure organization for
  open scholarship}. Quantitative Science Studies  \textbf{1}(1),  428--444 (02
  2020). \doi{10.1162/qss_a_00023}, \url{https://doi.org/10.1162/qss\_a\_00023}

\bibitem{skinner_values_2020}
Skinner, K., Lippincott, S.: Values and {Principles} {Framework} and
  {Assessment} {Checklist}. Tech. rep., Knowledge Futures Group (Jul 2020),
  \url{https://doi.org/10.21428/6ffd8432.5175bab1}

\bibitem{unesco_unesco_2021}
{UNESCO}: {UNESCO} {Recommendation} on {Open} {Science}. Programme and meeting
  document SC-PCB-SPP/2021/OS/UROS, {UNESCO} (2021),
  \url{https://unesdoc.unesco.org/ark:/48223/pf0000379949}

\bibitem{Wilkinson2016-te}
Wilkinson, M.D., Dumontier, M., Aalbersberg, I.J., Appleton, G., Axton, M.,
  Baak, A., Blomberg, N., Boiten, J.W., da~Silva~Santos, L.B., Bourne, P.E.,
  Bouwman, J., Brookes, A.J., Clark, T., Crosas, M., Dillo, I., Dumon, O.,
  Edmunds, S., Evelo, C.T., Finkers, R., Gonzalez-Beltran, A., Gray, A.J.G.,
  Groth, P., Goble, C., Grethe, J.S., Heringa, J., 't~Hoen, P.A.C., Hooft, R.,
  Kuhn, T., Kok, R., Kok, J., Lusher, S.J., Martone, M.E., Mons, A., Packer,
  A.L., Persson, B., Rocca-Serra, P., Roos, M., van Schaik, R., Sansone, S.A.,
  Schultes, E., Sengstag, T., Slater, T., Strawn, G., Swertz, M.A., Thompson,
  M., van~der Lei, J., van Mulligen, E., Velterop, J., Waagmeester, A.,
  Wittenburg, P., Wolstencroft, K., Zhao, J., Mons, B.: The {FAIR} guiding
  principles for scientific data management and stewardship. Sci. Data
  \textbf{3}(1),  160018 (Dec 2016). \doi{10.1038/sdata.2016.18}

\end{thebibliography}
\end{document}